\documentstyle[12pt]{article}
\begin{document}

\noindent {\small USC-98/HEP-B6\hfill \hfill hep-th/9812161}\newline
{\small \hfill }

{\vskip 2cm}

\begin{center}
{\Large {\bf Supersymmetric Two-Time Physics}}\medskip \footnote{%
This research was partially supported by the US. Department of Energy under
grant number DE-FG03-84ER40168.}

{\vskip 1cm}

{\bf I. Bars, C. Deliduman, D. Minic}

{\vskip 1cm}

{Department of Physics and Astronomy, University of Southern California}

{\ Los Angeles, CA 90089-0484, USA}

{\vskip 2cm}

{\bf Abstract}

{\vskip 0.5cm}
\end{center}

We construct an Sp(2,R) gauge invariant particle action which possesses
manifest space-time SO(d,2) symmetry, global supersymmetry and kappa
supersymmetry. The global and local supersymmetries are non-abelian
generalizations of Poincare type supersymmetries and are consistent with the
presence of two timelike dimensions. In particular, this action provides a
unified and explicit superparticle representation of the superconformal
groups OSp(N/4), SU(2,2/N) and OSp(8$^*$/N) which underlie various AdS/CFT
dualities in M/string theory. By making diverse Sp(2,R) gauge choices our
action reduces to diverse one-time physics systems, one of which is the
ordinary (one-time) massless superparticle with superconformal symmetry that
we discuss explicitly. We show how to generalize our approach to the case of
superalgebras, such as OSp(1/32), which do not have direct space-time
interpretations in terms of only zero branes, but may be realizable in the
presence of p-branes.

\newpage

\section{Introduction: what is two-time physics?}

Recently, a surprising connection between $SO(d,2)$ and point particle
physics has been uncovered in \cite{dualconf} and explored further in \cite
{dualH}-\cite{lifting}. The motivation for this work comes from the
algebraic approach to M-theory and its extensions to F-theory and S-theory.
The work in \cite{dualH}-\cite{lifting} has been called Two-Time Physics
(although it can be viewed as a reformulation of one-time physics) because
it involves the higher symmetry SO$\left( d,2\right) $ and because this
symmetry is realized{\it \ linearly} on a vector $X^M\left( \tau \right) $
with one extra time-like and one extra space-like dimension as compared to
the usual relativistic vector $x^\mu \left( \tau \right) $ in ordinary
spacetime in $\left( d-1\right) +1$ dimensions. Thanks to an Sp$\left(
2,R\right) $ gauge symmetry, which includes $\tau $-reparametrization as a
subgroup, the extra $1+1$ space-time dimensions can be related to Sp$\left(
2,R\right) $ gauge degrees of freedom at the cost of losing the manifest $%
SO(d,2)$ symmetry. This is analogous to relating $x^0\left( \tau \right) $
to $\tau $ by making a gauge choice at the cost of losing manifest SO$\left(
d-1,1\right) $ symmetry.

The most conservative way of understanding two-time physics is to view it as
a reformulation of ordinary one-time physics in $\left( d-1\right) +1$
dimensions that displays higher symmetries. Through this reformulation one
learns that the {\it action functional }for ordinary one-time physical
systems, for free or interacting particles, have a hidden space-time SO$%
\left( d,2\right) $ symmetry that is realized non-linearly. In the case of
the ordinary massless relativistic particle the symmetry is none other than
the usual conformal symmetry SO$\left( d,2\right) $ of massless systems.
However, the same global symmetry SO$\left( d,2\right) $ is also realized in
unusual ways in many other systems \cite{lifting}, all of which correspond
to different Sp$\left( 2\right) $ gauge choices. The other realizations of SO%
$\left( d,2\right) $ cannot be interpreted as conformal symmetries; they are
simply hidden symmetries of ordinary one-time systems that escaped detection
before (except for the well know SO$\left( 4,2\right) $ dynamical symmetry
of the hydrogen atom). Some specific examples that were treated in \cite
{dualconf}-\cite{lifting} are: massive non-relativistic particle, massive
relativistic particle, harmonic oscillator, hydrogen atom, particle in black
hole background, particle in an AdS$_D\times $S$^n$ background (with $D+n=d$%
), etc.. All of these have their analogs with spinning particles with spin $%
n/2$. In that case Sp$\left( 2\right) $ is replaced by OSp$\left( n/2\right) 
$ \cite{dualsusy}.

A remarkable connection between all these one-time systems is that they are
all connected to each other by Sp$\left( 2,R\right) $ or OSp$\left(
n/2\right) $ gauge transformations that play a role analogous to duality in
M-theory. In this context one may also make a connection to canonical
transformations in which the time is also transformed. The analogy to
duality in M-theory may be described as follows: usually one thinks of
various dual string theories as different corners of the moduli space of
M-theory that are connected to each other by duality (gauge)
transformations. Each one of the string theories is considered to be a
perturbative expansion of the full theory and somehow capable of capturing
the full M-theory by including all non-perturbative sectors. Thus, different
looking string theories are supposed to be just different starting points
for the same M-theory. In our case there is an analogous and possibly
stronger notion of duality. We called it ``duality'' to emphasize the
analogy for the benefit of the reader, but in the long term it probably
should be given a different name to avoid confusion. Namely, different
one-time physics models appear as different gauge choices of the same theory
and therefore they can be transformed into each other by gauge
transformations. These systems are unified together in one parent theory
that has both Sp$\left( 2,R\right) $ gauge symmetry (duality) and a higher
global symmetry SO$\left( d,2\right) $. Any version can be used to compute
the gauge invariant quantities of the parent theory. There is a single
action that describes all of them and thus unifies them. The unified theory
is the theory that we called Two-Time Physics for a zero brane. We have
indications that the same approach can be extended to p-branes.

The two previous paragraphs give a description of two-time physics from a
conservative point of view, involving only one-time physics, hidden
symmetries, and dualities that relate different systems. A more radical
point of view is offered by the manifestly Sp$\left( 2\right) $ (generally
OSp$\left( n/2\right) $) local gauge invariant and SO$\left( d,2\right) $
global invariant formulation. In this case the dynamical degrees of freedom
are the position-momentum Sp$\left( 2,R\right) $ doublet $X_i^M=\left(
X^M,P^M\right) $ which transforms linearly as an SO$\left( d,2\right) $
vector. The vector $X^M\left( \tau \right) $ is most directly interpreted as
the position of a particle in a space-time that includes two timelike
dimensions. The meaning of SO$\left( d,2\right) $ is simply the Lorentz
symmetry of the system in a flat spacetime with two times. This suggests a
very radical view of the meaning of ``time'' in one-time physics. To begin
with there are two timelike dimensions; which one is the physical time in
one-time physics? This is answered quite explicitly in the examples that
have been worked out in detail. Namely, the physical ``time'' in one-time
physics is not determined until one makes an Sp$\left( 2,R\right) $ gauge
choice. From the point of view of two-time physics the ``time'' of one-time
physics appears as a gauge artifact. One-time physics seems different from
one Sp$\left( 2,R\right) $ gauge choice to another because the choice of
time determines a corresponding one-time Hamiltonian (time evolution
generator for that choice of time) that looks different for every gauge
choice. However, in reality these very different looking one-time-physics
systems describe the same gauge invariant sector of the two-time physics
system.

What would be considered tests of two time physics? Whether one takes the
conservative or radical point of view some tests can provide evidence of the
unification of different one-time physics systems that occurs anyway. For
now we concentrate on theoretical tests and hope that we will produce some
experimental tests when we consider interacting two-time-physics systems.
The tests must concentrate on Sp$\left( 2,R\right) $ gauge invariants while
exploring the global SO$\left( d,2\right) $ symmetry. For two-time physics
the only gauge invariant observables are the generators of the global SO$%
\left( d,2\right) $ symmetry $L^{MN}=X^MP^N-X^NP^M$. All physical states or
other gauge invariant quantities can be computed fully covariantly without
choosing a gauge, or non-covariantly by choosing any gauge that defines a
one-time system. At the end, all gauge invariant quantities, including the
spectrum of the theory, are determined by the representation theory of SO$%
\left( d,2\right) $. It was demonstrated in \cite{dualconf}-\cite{lifting}
that indeed the quantum spectrum of the theory is the same in several
different gauges and that it corresponds to the unitary representation of SO$%
\left( d,2\right) $ with Casimir eigenvalues fully determined. In particular 
$C_2\left( SO\left( d,2\right) \right) =1-d^2/4$ for the spinless system and 
$C_2\left( SO\left( d,2\right) \right) =\frac 18\left( n-2\right) \left(
d+2\right) \left( d+n-2\right) $ for the spinning system with spin $n/2$.
The difference between Hilbert spaces of diverse but dual one-time models is
only in the choice of basis labelled by different quantum numbers. These
bases are related to each other by unitary transformations (duality at the
quantum level) within the same SO$\left( d,2\right) $ representation. This
test was verified in several different gauges. For example the H-atom and
free massless particle (and all other cases)\ correspond to the same SO$%
\left( d,2\right) $ representation with the same Casimir eigenvalues. This
was not known before the advent of two-time physics. This is a test at the
quantum level. A similar test at the classical level is trivial because all
Casimir eigenvalues vanish due to constraints; the test becomes non-trivial
at the quantum level due to quantum ordering.

Whether one takes the conservative or the radical point of view, the unified
approach is also useful in uncovering and understanding the symmetries of
one-time physics. The fact that previously unknown symmetries have been
found in very familiar systems is a success of the approach \cite{lifting}.
The hidden symmetries arise naturally in fixed gauges as the natural SO$%
\left( d,2\right) $ symmetry of the original theory. Much of this symmetry
remains hidden in the one-time physics formulation. For example, it was
discovered that the field theory that describes a free scalar particle in an
AdS$_D\times S^n$ background has SO$\left( D+n,2\right) $ symmetry, which is
larger than the previously known Killing symmetry SO$\left( D-1,2\right)
\times SO\left( n+1\right) $. Thus, for AdS$_5\times $S$^5$ \cite{maldacena}
the symmetry is SO$\left( 10,2\right) $ not just SO$\left( 4,2\right) \times
SO\left( 6\right) $. For the symmetry to be valid the particle must have a
quantized mass as shown in \cite{lifting}.

In this paper we generalize the two-time zero brane system to spacetime
supersymmetry. We construct an Sp(2,R) gauge invariant particle action which
possesses a manifest space-time superconformal symmetry. In particular, we
demonstrate that the superconformal groups $OSp(N/4)$, $SU(2,2/N)$ and $%
OSp(8^{*}/N)$ which underlie various $AdS/CFT$ dualities in M/string theory
are given unified and explicit superparticle representations. Our approach
can be generalized for the case of superconformal algebras, such as
OSp(1/32), which do not have direct space-time interpretation in terms of
only zero branes, but could be realized in the presence of p-branes.

As in \cite{dualconf} we emphasize that the $Sp(2,R)$ symmetry that is
gauged may also be viewed via $Sp(2,R)=SO(1,2)$ as the conformal group in $%
0+1$ dimensions. From this point of view, many $0+1$ quantum gravity systems
(relativistic massless and massive particle, non-relativistic massless and
massive particle, particle moving in the $AdS_D\times S^n$ background etc.)
can be viewed as different gauge choices of the same $0+1$ conformal quantum
gravity theory.

One gauge choice that we will investigate in detail corresponds to the
standard superparticle in one-time physics. It has been known for a long
time \cite{schwarz} that the massless superparticle has actually
non-linearly realized hidden superconformal symmetry, extending the
conformal symmetry SO$\left( d,2\right) $ of massless systems for $d=3,4$,6.
In our approach the hidden conformal supersymmetries of the superparticle
action $OSp(N/4)$, $SU(2,2/N)$, $OSp(8^{*}/N)$ are just the explicit
supersymmetries in the two-time physics formulation. This is another example
of the utility of our approach, now in a supersymmetric setting. This also
gives a two-time meaning to the superparticle system as well as to
supersymmetry itself.

Finally, there are other possible gauge choices of our Lagrangian, which
would provide explicit unusual realizations of the various space-time
supersymmetries $OSp(N/4)$, $SU(2,2/N)$, $OSp(8^{*}/2N)$. Guided by the
analogy with the bosonic case, we list just a few: superparticles moving in
the background of $AdS_D\times S^n$, supersymmetric H-atom, supersymmetric
particle moving in the black hole background, etc.. The case of
supersymmetric particle in $AdS_D\times S^n$ would be of special current
interest. We predict, following the bosonic case investigated in \cite
{lifting}, that the field theory that describes the particle in the
supersymmetric $AdS_5\times S^5$ background \cite{maldacena} has OSp$\left(
8^{*}/8\right) $ supersymmetry and not just the symmetry of its subgroup SU$%
\left( 2,2/4\right) $; similarly, the supersymmetric $AdS_3\times S^3$
background \cite{maldstrom} has OSp$\left( 4/4\right) $ supersymmetry and
not just the symmetry of its subgroup SU$\left( 1,1/2\right) $ .

\section{Lagrangian: why two time-like dimensions?}

The dynamical variables are the bosons $X_i^M\left( \tau \right) $ and
fermions $\Theta ^{\alpha a}\left( \tau \right) $. The $X_i^M$ form an Sp$%
\left( 2,R\right) $ doublet, $i=1,2$, while the $\Theta ^{\alpha a}\left(
\tau \right) $ are Sp$\left( 2,R\right) $ singlets. The $X_i^M\left( \tau
\right) $ are position-momentum vectors in $d+2$ dimensions, $M=1,\cdots
,d+2 $. The $\Theta ^{\alpha a}\left( \tau \right) $ are spinors of SO$%
\left( d,2\right) $, $\alpha =1,\cdots ,s\left( d\right) $, and there are $N$
of them $a=1,\cdots ,N$. To construct a covariant derivative $D_\tau X_i^M$
as an Sp$\left( 2,R\right) $ doublet we use the Sp$\left( 2,R\right) $ gauge
field $A^{ij}=A^{ji}$ (a triplet) as in \cite{dualconf} (there is only the
time component $A_\tau ^{ij}=A^{ij}$ since we are on the worldline) 
\begin{equation}
D_\tau X_i^M=\partial _\tau X_i^M-\varepsilon _{ij}A^{jk}X_k^M-\Omega
^{MN}X_{iN}.  \label{covD}
\end{equation}
The new $\Omega ^{MN}$ (an Sp$\left( 2,R\right) $ singlet) will play a role
for supersymmetry as explained below. The antisymmetric $\varepsilon _{ij}$
is the Sp$\left( 2,R\right) $ metric that is used to lower or raise indices.

The Sp$\left( 2,R\right) $ gauge invariant Lagrangian is 
\begin{eqnarray}
L &=&\frac 12D_\tau X_i^MX_j^N\varepsilon ^{ij}\eta _{MN}\,,
\label{invlagrangian} \\
&=&\partial _\tau X_1\cdot X_2-\frac 12A^{ij}X_i\cdot X_j-\frac 12\Omega
^{MN}L_{MN}\,\,.
\end{eqnarray}
The first term in the Lagrangian is simplified by doing an integration by
parts so that one may identify $X_1^M=X^M$ as position and $%
X_2^M=P^M=\partial L/\partial \left( \partial _\tau X_{1M}\right) $ as
momentum. Then the gauge invariant 
\begin{equation}
L_{MN}=X_i^MX_j^N\varepsilon ^{ij}=X^MP^N-X^NP^M  \label{orbital}
\end{equation}
is the orbital angular momentum that generates SO$\left( d,2\right) $ on the
bosons $\left( X^M,P^M\right) $.

We recall why one {\it must have two timelike dimensions} as in \cite
{dualconf}. The equations of motion of the gauge field $A^{ij}$ demand the
constraints (for related work see \cite{marnelius}\cite{siegel}) 
\begin{equation}
X_i\cdot X_j=0,\quad or\quad X^2=P^2=X\cdot P=0,  \label{xixj}
\end{equation}
which require that the two light-like vectors $\left( X^M,P^M\right) $ are
orthogonal. In a spacetime with only one timelike dimension, $X^M$ and $P^M$
must be parallel lightlike vectors with zero angular momentum $L^{MN}=0$.
Since $L^{MN}$ is the only gauge invariant observable, this becomes a
trivial theory if spacetime contains only one timelike dimension. To have
solutions of the constraints with non-trivial angular momentum $L_{MN}$ one
must admit a spacetime metric $\eta _{MN}$ that has two time-like
dimensions. The Sp$\left( 2,R\right) $ gauge symmetry is just enough to
remove all the ghosts introduced by the two timelike dimensions and provide
a unitary theory. One cannot have more than two time-like dimensions because
the theory would be non-unitary due to ghosts that cannot be removed.

There are ways of having more than two timelike dimensions in more
complicated multi-particle (or multi-string) systems \cite{multi}, but not
in the system of one zero brane.

\section{Fermions and spin connection}

In the covariant derivative (\ref{covD}) we have included a new term
involving $\Omega ^{MN}\left( \Theta ,\partial _\tau \Theta \right) $ which
is constructed from the fermions $\Theta ^{\alpha a}\left( \tau \right) $
and their derivative. The role of this term is to insure a covariant
derivative $D_\tau X_i^M$ under {\it global} supersymmetry. For this to be
possible, we will see that $\Omega ^{MN}$ transforms like a spin connection
under field dependent {\it local} SO$\left( d,2\right) $ transformations
induced by global supersymmetry transformations. $\Omega ^{MN}$ is given as
a supertrace 
\begin{equation}
\Omega ^{MN}=Str\left( \left( 
\begin{array}{cc}
\Gamma ^{MN} & 0 \\ 
0 & 0
\end{array}
\right) \left( \partial _\tau t\right) t^{-1}\right) ,\quad t\left( \Theta
\right) \in G/H  \label{Omn}
\end{equation}
where $t\left( \Theta \right) $ is an element of the coset $G/H$
parametrized by the fermions $\Theta ^{\alpha a}.$ An explicit
parametrization of $t\left( \Theta \right) $ and $\Omega ^{MN}\left( \Theta
\right) $ are given below, but only the general properties of cosets rather
than explicit expressions are essential for most of the discussion. The
coset $G/H$ is given by one of the following three cases\footnote{%
The reason for considering only these supergroups is that the spacetime
subgroup is precisely SO$\left( d,2\right) $ for $d=3,4,6$ respectively. For
these cases the orbital part of SO$\left( d,2\right) $, i.e. $L^{MN}$, can
be constructed from dynamical variables $X_i^M\left( \tau \right) $ that are
zero branes, as in $\left( \ref{orbital}\right) $. For other supergroups,
such as OSp$\left( 1/32\right) $, additional dynamical bosonic variables
that are possibly related to p-branes must be included. Except for comments
on the general construction of the Lagrangian found in the discussion
section, we leave the study of such cases to future research.} 
\begin{eqnarray}
&&OSp\left( N/4\right) /\left[ SO\left( N\right) \times Sp\left( 4,R\right) 
\right] ,\quad \\
&&SU\left( 2,2/N\right) /\left[ SU\left( N\right) \times SU\left( 2,2\right) 
\right] , \\
&&OSp\left( 8^{*}/N\right) /\left[ Sp\left( N\right) \times Spin\left(
8^{*}\right) \right] ,\quad N=even
\end{eqnarray}
Notice that precisely these supergroups occur naturally in the AdS/CFT
correspondence in M/IIB string theory. The subgroup $H$ includes the
spacetime Lorentz group SO$\left( d,2\right) $, with two timelike dimensions
for $d+2=5,6,8$, (or the conformal group in d dimensions), namely 
\begin{equation}
Sp\left( 4,R\right) =SO\left( 3,2\right) ,\quad SU\left( 2,2\right)
=SO\left( 4,2\right) ,\quad Spin\left( 8^{*}\right) =SO\left( 6,2\right) .
\end{equation}
The number of supersymmetries is $N$ (real)$,N$ (complex)$,N$ (complex and
even) respectively in the three cases. The generators of SO$\left(
d,2\right) $ acting on the spinors ($\alpha $ index) is constructed from
gamma matrices in the form $\Gamma ^{MN}=\frac 12\left[ \Gamma ^M,\Gamma ^N%
\right] $, and it is embedded in the spacetime block of the matrix
representation of the groups above, as seen in (\ref{Omn}).

$t\left( \Theta \right) =\exp \left( \vartheta ^{\alpha a}Q_{\alpha
a}\right) $ is given by exponentiation the off-diagonal coset generators $%
Q_{\alpha a}$ (represented by the $s\left( d\right) \times N$ rectangular
blocks) combined with the parameters $\vartheta ^{\alpha a}$%
\begin{equation}
t\left( \Theta \right) =\exp \left( 
\begin{array}{cc}
0 & \vartheta \\ 
-\tilde{\vartheta} & 0
\end{array}
\right) =\left( 
\begin{array}{ll}
\frac 1{\sqrt{1+\Theta \tilde{\Theta}}} & \Theta \frac 1{\sqrt{1+\tilde{%
\Theta}\Theta }} \\ 
-\frac 1{\sqrt{1+\tilde{\Theta}\Theta }}\tilde{\Theta} & \frac 1{\sqrt{1+%
\tilde{\Theta}\Theta }}
\end{array}
\right) \,\,  \label{tmatrix}
\end{equation}
where $\Theta =\vartheta \frac{\tan \sqrt{\tilde{\vartheta}\vartheta }}{%
\sqrt{\tilde{\vartheta}\vartheta }}.$ The square roots or $\tan \sqrt{\tilde{%
\vartheta}\vartheta }$ are understood in the sense of infinite series. The $%
\tilde{\vartheta}_{\alpha a},\tilde{\Theta}_{\alpha a}$ are obtained by
transposition or hermitian conjugation combined with multiplication by
appropriate metrics or charge conjugation matrices $C$ for the spinors, 
\begin{equation}
\tilde{\Theta}_{a\alpha }=\left( \Theta ^TC\right) _{a\alpha
},\,\,\,\,\left( \Theta ^{\dagger }C\right) _{a\alpha },\,\,\,\,\left(
K\Theta ^TC\right) _{a\alpha },
\end{equation}
respectively for the three cases$.$ Here the unit matrix $\delta _{ab}$ is
the metric for SO$\left( N\right) $, the unit matrix $\delta _a^b=\delta _{a%
\dot{b}}$ is the metric for SU$\left( N\right) $, and the antisymmetric
matrix $K_{ab}$ is the metric for Sp$\left( N\right) $.

According to a general theorem, an element $g$ of any (super)group can
always be decomposed into the product of an element of the subgroup $h$
times an element of the coset $t$, thus $g=ht$. Following this, a basic
relation that we will use in our case is ($t_1t_2$ is a group element $g$) 
\begin{equation}
t\left( \Theta _1\right) t\left( \Theta _2\right) =h\left( \Theta _1,\Theta
_2\right) \,t\left( \Theta _{12}\right)  \label{ttht}
\end{equation}
where both $h\left( \Theta _1,\Theta _2\right) $ and $t\left( \Theta
_{12}\right) $ are functions of $\Theta _1,\Theta _2$. Explicit forms for $%
h\left( \Theta _1,\Theta _2\right) $ and $t\left( \Theta _{12}\right) $ are
given in \cite{barsgunay} and in the appendix, but these details are not
needed in most of our discussion. In our case $t\left( \Theta _{12}\right) $
depends on $\Theta _{12}$ which takes values in the off-diagonal blocks and
has the form (\ref{tmatrix}) while $h\left( \Theta _1,\Theta _2\right) $
takes values in the diagonal blocks with the upper block being the spinor
representation of SO$\left( d,2\right) $ and the lower block the fundamental
representation of SO$\left( N\right) $, SU$\left( N\right) $, Sp$\left(
N\right) $ respectively.

\section{Global supersymmetry}

The global supersymmetry transformation on the fermions $\Theta \rightarrow
\Theta _\varepsilon $ is defined by using the relation (\ref{ttht}) for $%
\Theta _1=\Theta $, $\Theta _2=\varepsilon $, $\Theta _{12}=\Theta
_\varepsilon $. Since $\Theta $ appears only in the form $t\left( \Theta
\right) $ the SUSY transformation may be written as the following product of
super matrices 
\begin{equation}
t\left( \Theta \right) \rightarrow t\left( \Theta _\varepsilon \right)
=h^{-1}\left( \Theta ,\varepsilon \right) \,t\left( \Theta \right) \,t\left(
\varepsilon \right) .  \label{globalsusy}
\end{equation}
Inserting this in (\ref{Omn}) we find 
\begin{eqnarray}
&&\Omega ^{MN}\left( \Theta _\varepsilon \right)  \nonumber \\
&=&Str\left( \left( 
\begin{array}{cc}
\Gamma ^{MN} & 0 \\ 
0 & 0
\end{array}
\right) \partial _\tau \left[ h^{-1}\left( \Theta ,\varepsilon \right)
t\left( \Theta \right) t\left( \varepsilon \right) \right] \,t^{-1}\left(
\varepsilon \right) t^{-1}\left( \Theta \right) h\left( \Theta ,\varepsilon
\right) \right) \\
&=&Str\left( \left( 
\begin{array}{cc}
\Gamma ^{MN} & 0 \\ 
0 & 0
\end{array}
\right) h^{-1}\left( \left( \partial _\tau t\right) t^{-1}-\partial _\tau
\right) h\right) \\
&=&\left[ \Lambda ^{-1}\left( \Omega \left( \Theta \right) -\partial _\tau
\right) \Lambda \right] ^{MN}  \label{Oprime}
\end{eqnarray}
where $\Lambda _{\,\,\,N}^M\left( \Theta ,\varepsilon \right) $ is a field
dependent local SO$\left( d,2\right) $ transformation and $\Omega
^{MN}\left( \Theta \right) $ transforms like a spin connection. In arriving
at this result we used that $t\left( \varepsilon \right) $ is independent of 
$\tau $, and noted that $\Lambda _{\,\,\,N}^M\left( \Theta ,\varepsilon
\right) $ is the vector representation of SO$\left( d,2\right) $ whose
spinor representation corresponds to the upper block in $h\left( \Theta
,\varepsilon \right) $. Note that $\Omega ^{MN}$ is invariant under the
internal symmetry transformations contained in the lower block of $h\left(
\Theta ,\varepsilon \right) $.

We now turn to the transformations of $X_{i}^{M}$ and the covariant
derivative $D_{\tau }X_{i}^{M}$ in (\ref{covD}). We see that, if we take the
supersymmetry transformation for $X_{i}^{M}$ to be the induced local Lorentz
transformation 
\begin{equation}
X_{i}^{M}\rightarrow X_{i\varepsilon }^{M}=\left( \Lambda ^{-1}\right)
_{\,\,\,N}^{M}\left( \Theta ,\varepsilon \right) X_{i}^{N}\,\,,
\end{equation}
then the covariant derivative also transforms covariantly under the {\it %
local} Lorentz transformation since $\Omega ^{MN}$ transforms like a spin
connection 
\begin{equation}
D_{\tau }X_{i}^{M}\rightarrow \left( \Lambda ^{-1}\right)
_{\,\,\,N}^{M}\left( \Theta ,\varepsilon \right) \left( D_{\tau
}X_{i}^{N}\right) \,.
\end{equation}
Since the Lagrangian is already invariant under Lorentz transformations SO$%
\left( d,2\right) $, it is also invariant under the local transformation $%
\Lambda _{\,\,\,N}^{M}\left( \Theta ,\varepsilon \right) $. Hence it is
invariant under the global supersymmetry transformation generated by $%
t\left( \varepsilon \right) $ as given by eqs. (\ref{tmatrix}) and (\ref
{globalsusy}).

The global symmetry transformations with the subgroup $H$ including SO$%
\left( d,2\right) $ and SO$\left( N\right) $, SU$\left( N\right) $, Sp$%
\left( N\right) $ proceeds in the same way. Instead of eq.(\ref{ttht})
consider the product $t\left( \Theta \right) h$ where is global $h\in H$.
Using the general theorem we can write $t\left( \Theta \right) h=ht\left(
\Theta _h\right) $ where $\Theta _h$ is the transformed $\Theta $ so that 
\begin{equation}
t\left( \Theta \right) \rightarrow t\left( \Theta _h\right) =h^{-1}\,t\left(
\Theta \right) \,h
\end{equation}
In the present case the same global $h$ must appear on both sides of $%
t\left( \Theta \right) $ since this is the only way for $t\left( \Theta
\right) $ to maintain the matrix form in (\ref{tmatrix}). Following the same
steps we learn that $\Omega ^{MN}$ transforms like (\ref{Oprime}) but with a
global $\Lambda $. Therefore $X_i^M$ also transforms with a global $\Lambda $
and the Lagrangian is invariant. Note that only $\Theta $ transforms under
the global internal symmetry. Also $A^{ij}$ does not transform under any of
the global symmetries.

This establishes that the global symmetries of the Lagrangian are all the
transformations of the supergroup $G=$OSp$\left( N/4\right) $, SU$\left(
2,2/N\right) $, OSp$\left( 8^{*}/N\right) $ respectively for the various
dimensions $d+2=5,6,8$. All global transformations originate from the
multiplication of $t\left( \Theta \right) $ from the right, $t\left( \Theta
\right) g$, where $g$ is a global group element $g\in G$. By construction it
is seen that these transformations close to form the respective supergroups.
The conserved generators are easily derived by using Noether's theorem, we
find 
\begin{eqnarray}
\Psi _{\alpha a} &=&\left( \frac 1{\sqrt{1+\Theta \tilde{\Theta}}}\left(
\frac 12\Gamma ^{KL}L_{KL}\right) \,\Theta \frac 1{\sqrt{1+\tilde{\Theta}%
\Theta }}\right) _{\alpha a}  \label{psia} \\
J^{MN} &=&Tr\left( \Gamma ^{MN}\frac 1{\sqrt{1+\Theta \tilde{\Theta}}}\left(
\frac 12\Gamma ^{KL}L_{KL}\right) \frac 1{\sqrt{1+\Theta \tilde{\Theta}}%
}\right)  \label{JMN} \\
T^A &=&\frac 12\left( \tilde{\Theta}_\alpha t^A\Psi ^\alpha +h.c.\right) ,
\label{VA}
\end{eqnarray}
where $\left( t^A\right) _{\,\,b}^a$ are the hermitian matrix
representations for the generators of the internal group SO$\left( N\right) $%
, SU$\left( N\right) $, or Sp$\left( N\right) $ in the fundamental
representation.

\section{Local kappa symmetry}

The local kappa supersymmetry transformation on the fermions $\Theta
\rightarrow \Theta _\kappa $ is defined by using the relation (\ref{ttht})
for $\Theta _1=\kappa $, $\Theta _2=\Theta $, $\Theta _{12}=\Theta _\kappa $%
. We will see below that the local spinors $\kappa ^{\alpha a}$ must be
constructed from a specific combination of $X_i^M$ and two independent local
spinors $\kappa ^{i\alpha a}\left( \tau \right) $, $i=1,2,$ as follows 
\begin{equation}
\kappa ^{\alpha a}=\left( \not{X}_i\kappa ^i\right) ^{\alpha a}.
\end{equation}
This transformation is realized by the product of super matrices 
\begin{equation}
t\left( \Theta \right) \rightarrow t\left( \Theta _\kappa \right)
=h^{-1}\left( \kappa ,\Theta \right) \,t\left( \kappa \right) \,t\left(
\Theta \right) \,.
\end{equation}
Comparing to the global transformation (\ref{globalsusy}) we emphasize that
the global $t\left( \varepsilon \right) $ is a right multiplication while
the local $t\left( \kappa \right) $ is a left multiplication\footnote{%
The group theoretical origin of kappa supersymmetry for the ordinary
superparticle or super p-branes has remained obscure. In our treatment it is
seen to originate from a group transformation similar to the global
supersymmetry, except that it acts on the other side of the coset element $%
t\left( \Theta \right) $. Indeed the ordinary superparticle can be
reformulated in precisely the same way. For previous approaches to kappa
supersymmetry see \cite{sorokin}\cite{tonin}\cite{howe}.} with $t\left(
\Theta \right) $. We also emphasize that the order of the arguments in $%
h\left( \kappa ,\Theta \right) $ and $h\left( \Theta ,\varepsilon \right) $
are interchanged because of the same reason. Inserting this form in (\ref
{Omn}) we find 
\begin{eqnarray}
&&\Omega ^{MN}\left( \Theta _\kappa \right)  \nonumber \\
&=&Str\left( \left( 
\begin{array}{cc}
\Gamma ^{MN} & 0 \\ 
0 & 0
\end{array}
\right) \partial _\tau \left[ h^{-1}\left( \kappa ,\Theta \right) g\left(
\kappa ,\Theta \right) \right] \,g^{-1}\left( \kappa ,\Theta \right)
\,h\left( \kappa ,\Theta \right) \right) \\
&=&Str\left( \left( 
\begin{array}{cc}
\Gamma ^{MN} & 0 \\ 
0 & 0
\end{array}
\right) h^{-1}\left[ \left( \partial _\tau g\right) g^{-1}-\partial _\tau 
\right] h\right)
\end{eqnarray}
where we have defined 
\begin{equation}
g\left( \kappa ,\Theta \right) =t\left( \kappa \right) \,t\left( \Theta
\right) .
\end{equation}
As in (\ref{Oprime}) we can rewrite 
\begin{equation}
\Omega ^{MN}\left( \Theta _\kappa \right) =\left[ \Lambda ^{-1}\left( \Omega
_g-\partial _\tau \right) \Lambda \right] ^{MN},  \label{Okappa}
\end{equation}
where $\Lambda _{\,\,\,N}^M\left( \kappa ,\Theta \right) $ is a field
dependent local SO$\left( d,2\right) $ transformation similar to the
previous case, except for interchanging the orders of the two fermions and
substituting $\kappa $ instead of $\varepsilon $. Furthermore, $\Omega
_g^{MN}$ is not $\Omega ^{MN}\left( \Theta \right) $, since it is
constructed from $g\left( \kappa ,\Theta \right) $ not from $t\left( \Theta
\right) $ as in (\ref{Omn}).

We now turn to the kappa transformations of $X_i^M$. We see that, if we take
the supersymmetry transformation for $X_i^M$ to be the induced local Lorentz
transformation 
\begin{equation}
X_i^M\rightarrow X_{i\kappa }^M=\left( \Lambda ^{-1}\right)
_{\,\,\,N}^M\left( \kappa ,\Theta \right) X_i^N\,\,,
\end{equation}
then the covariant derivative transforms into 
\begin{equation}
D_\tau X_i^M\rightarrow \left( \Lambda ^{-1}\right) _{\,\,\,N}^M\left(
\kappa ,\Theta \right) \left( \partial _\tau X_i^N-\varepsilon _{ij}A_\kappa
^{jk}X_k^N-\Omega _g^{NK}X_{iK}\right) \,.
\end{equation}
The right hand side is not the covariant derivative since it contains $%
\Omega _g^{NK}$ instead of $\Omega ^{NK}\left( \Theta \right) $, and the
transformed $A_\kappa ^{jk}$ instead of $A^{ij}$. Inserting these
transformations into the Lorentz invariant Lagrangian we see that $\Lambda
_{\,\,\,N}^M\left( \kappa ,\Theta \right) $ drops out, and we find the
result 
\begin{equation}
L\left( X_\kappa ,\Theta _\kappa ,A_\kappa \right) =L\left( X,\Theta
,A\right) -\frac 12L_{MN}\left[ \Omega _g^{MN}-\Omega ^{MN}\left( \Theta
\right) \right] -\frac 12\left( A_\kappa ^{ij}-A^{ij}\right) X_i\cdot X_j.
\end{equation}
The condition for kappa invariance is the vanishing of the last two terms
which may be written in the form 
\begin{equation}
Str\left( \left( 
\begin{array}{cc}
L & 0 \\ 
0 & 0
\end{array}
\right) \left[ \left( \partial _\tau g\right) g^{-1}-\left( \partial _\tau
t\right) t^{-1}\right] \right) =-\frac 12\left( A_\kappa ^{ij}-A^{ij}\right)
X_i\cdot X_j  \label{ksymmetry}
\end{equation}
with $L=\frac 12L_{MN}\Gamma ^{MN}$. We can examine this equation for
infinitesimal $\kappa $ by using 
\begin{eqnarray}
g\left( \kappa ,\Theta \right) &=&\left( 1+\left( 
\begin{array}{ll}
0 & \kappa \\ 
\tilde{\kappa} & 0
\end{array}
\right) +\cdots \right) t\left( \Theta \right) \\
\left( \partial _\tau g\right) g^{-1}-\left( \partial _\tau t\right) t^{-1}
&=&\left( 
\begin{array}{ll}
0 & \partial _\tau \kappa \\ 
\partial _\tau \tilde{\kappa} & 0
\end{array}
\right) +\left[ \left( 
\begin{array}{ll}
0 & \kappa \\ 
\tilde{\kappa} & 0
\end{array}
\right) ,\left( \partial _\tau t\right) t^{-1}\right] +\cdots
\end{eqnarray}
Therefore (\ref{ksymmetry}) becomes 
\begin{equation}
Str\left( \left( 
\begin{array}{ll}
0 & L\kappa \\ 
-\tilde{\kappa}L & 0
\end{array}
\right) \left( \partial _\tau t\right) t^{-1}\right) +\frac 12\delta _\kappa
A^{ij}X_i\cdot X_j=0.
\end{equation}
This will vanish only if $L\kappa $ is proportional to $X_i\cdot X_j$ since
then $\delta _\kappa A^{ij}$ can be chosen to cancel its coefficient.
Fortunately this is easily arranged by taking $\kappa ^{\alpha a}=\left( 
\not{X}_i\kappa ^i\right) ^{\alpha a}$ where $\kappa ^{i\alpha a}\left( \tau
\right) $ are two independent local fermions $i=1,2$. Indeed, we then find 
\begin{equation}
L\kappa =\frac 12\varepsilon ^{kj}X_k^MX_j^NX_i^K\left( \Gamma _{MN}\Gamma
_K\kappa ^i\right) =-X_i\cdot X_j\left( \not{X}_k\kappa ^i\right)
\varepsilon ^{jk}.  \label{manipulate}
\end{equation}
A three gamma term $\Gamma _{MNK}$ that could have appeared on the
right-hand side vanishes because it imposes antisymmetry in $M,N,K$ which is
impossible to have with only two independent vectors ($X^M$, $P^M$).
Therefore the kappa transformation for $A^{ij}$ must be 
\begin{equation}
\delta _\kappa A^{ij}=\frac 12Str\left( \left( 
\begin{array}{cc}
0 & \left( \not{X}_k\kappa ^{(i}\right) \varepsilon ^{j)k} \\ 
-\left( \tilde{\kappa}^{(i}\not{X}_k\varepsilon ^{j)k}\right) & 0
\end{array}
\right) \partial _\tau t\left( \Theta \right) t^{-1}\left( \Theta \right)
\right) .
\end{equation}
With the chosen transformations for $\Theta ^{\alpha a},X_i^M,A^{ij}$ the
Lagrangian is invariant under the kappa supersymmetry with the local
parameters $\kappa ^{i\alpha a}\left( \tau \right) $ which form an Sp$\left(
2,R\right) $ doublet $i=1,2$. Due to the local symmetry there is a
corresponding constraint on the canonical degrees of freedom that takes the
form 
\begin{equation}
\not{X}_i\sqrt{1+\Theta \tilde{\Theta}}\Psi =0.
\end{equation}
This is easily verified by using (\ref{psia}), the manipulations of (\ref
{manipulate}), and the constraints (\ref{xixj}).

\section{Gauge fixing to superparticle}

Consider the gamma matrices $\gamma ^\mu $ appropriate for SO$\left(
d-1,1\right) $ and construct the gamma matrices appropriate for SO$\left(
d,2\right) $ by taking direct products with $\tau _3$ and $\tau ^{\pm
}=\frac 1{\sqrt{2}}\left( \tau _1\pm i\tau _2\right) $, as follows 
\begin{eqnarray}
M &=&\left( \quad \pm ^{\prime }\quad ,\quad \quad \mu \quad \right) \\
\Gamma ^M &=&\left( \pm \tau ^{\pm }\times 1,\quad \tau _3\times \gamma ^\mu
\right) \\
\left\{ \Gamma ^M,\Gamma ^N\right\} &=&2\eta ^{MN},\quad \eta ^{+^{\prime
}-^{\prime }}=-1,\quad \eta ^{\mu \nu }=Minkowski
\end{eqnarray}
For our purposes we need to construct $\Gamma ^{MN}=\frac 12\left[ \Gamma
^M,\Gamma ^N\right] $, which takes the form 
\begin{equation}
\Gamma ^{MN}:\,\,\Gamma ^{+^{\prime }-^{\prime }}=-\tau _3\times 1,\quad
\Gamma ^{\pm ^{\prime }\mu }=-\tau ^{\pm }\times \gamma ^\mu ,\quad \Gamma
^{\mu \nu }=1\times \gamma ^{\mu \nu }.
\end{equation}

We now fix two of the Sp$\left( 2,R\right) $ and half of the kappa gauge
symmetries as follows 
\begin{equation}
X^{+^{\prime }}=1,\quad P^{+^{\prime }}=0,\quad \Gamma ^{+^{\prime }}\Theta
^{a}=0.
\end{equation}
After solving the two constraints $X^{2}=X\cdot P=0$, the gauge fixed forms
of $X,P$ are 
\begin{eqnarray}
M &=&\left( +^{\prime },\quad -^{\prime },\quad \mu \right)  \\
X^{M} &=&\left( 1,\quad \frac{x^{2}}{2},\quad x^{\mu }\right)  \\
P^{M} &=&\left( 0,\quad x\cdot p,\quad p^{\mu }\right) .
\end{eqnarray}
Using $\Gamma ^{+^{\prime }}=\tau ^{+}\times 1$ and $C=\tau _{1}\times c$,
we can also write explicitly the gauge fixed form of $\Theta ^{a},\tilde{%
\Theta}_{a}$%
\begin{equation}
\Theta ^{a}=\left( 
\begin{array}{l}
\theta ^{a} \\ 
0
\end{array}
\right) ,\quad \tilde{\Theta}_{a}=\left( 
\begin{array}{ll}
0 & \tilde{\theta}_{a}
\end{array}
\right) 
\end{equation}
Before gauge fixing $\Omega ^{MN}$ is computed by using (\ref{Omn}) and (\ref
{tmatrix})

\begin{equation}
\Omega ^{MN}=Tr\left( \Gamma ^{MN}\frac 1{\sqrt{1+\Theta \tilde{\Theta}}%
}\left( \partial _\tau \Theta \,\tilde{\Theta}\frac 1{\sqrt{1+\Theta \tilde{%
\Theta}}}-\partial _\tau \sqrt{1+\Theta \tilde{\Theta}}\right) \right)
\end{equation}
Then for the gauge fixed $\Theta $ we obtain $\left( \tilde{\Theta}\Theta
\right) _a^{\,\,\,b}$ = $\left( \tilde{\Theta}\partial _\tau \Theta \right)
_a^{\,\,\,b}$ = $0$, etc.. Therefore, by the series expansion of the square
roots, the expression for $\Omega ^{MN}$ simplifies to $\Omega ^{MN}=-\tilde{%
\Theta}_a\Gamma ^{MN}\partial _\tau \Theta ^a$, although still all of its
components are zero except for the one that contains $\Gamma ^{-^{\prime
}\mu }=-\tau ^{-}\times \gamma ^\mu $. Thus, 
\begin{eqnarray}
\Omega ^{-^{\prime }\mu } &=&-\tilde{\Theta}_a\Gamma ^{-^{\prime }\mu
}\partial _\tau \Theta ^a=-\tilde{\theta}_a\gamma ^\mu \partial _\tau \theta
^a, \\
\frac 12\Omega ^{MN}L_{MN} &=&\Omega ^{-^{\prime }\mu }L_{-^{\prime }\mu }=-%
\tilde{\theta}_a\gamma ^\mu \partial _\tau \theta ^ap_\mu ,
\end{eqnarray}
where we have used $L_{-^{\prime }\mu }=\eta _{-^{\prime }+^{\prime }}\,\eta
_{\mu \nu }\,L^{+^{\prime }\nu }=-p_\mu $. Therefore, in this gauge, the
invariant Lagrangian (\ref{invlagrangian}) collapses to the superparticle
Lagrangian 
\begin{eqnarray}
L &=&\dot{x}\cdot p-\frac 12A^{22}p^2+\tilde{\theta}_a\not{p}\partial _\tau
\theta ^a, \\
&=&\frac 1{2A^{22}}\left( \dot{x}^\mu +\tilde{\theta}_a\gamma ^\mu \partial
_\tau \theta ^a\right) ^2,
\end{eqnarray}
where the last form is obtained by integrating out $p^\mu $.

The superparticle Lagrangian is well known to have local symmetries that
correspond to $\tau $ reparametrization and kappa supersymmetry. Because of $%
\tau $ reparametrization (part of Sp$\left( 2,R\right) $ that has not been
gauge fixed) there remains one bosonic constraint $P^2=p^2=0$, and because
of the remaining kappa supersymmetry there remains a fermionic constraint $%
\not{p}Q=0$ where $Q$ is the global supersymmetry generator $Q=\not{p}\theta 
$.

It has been known for some time \cite{schwarz} that this Lagrangian also has
hidden superconformal symmetries. In our approach the presence of the hidden
superconformal symmetry is a natural consequence of the manifestly
supersymmetric Lagrangian we started from, which had global OSp$\left(
N/4\right) $, $SU\left( 2,2/N\right) $ or OSp$\left( 8^{*}/N\right) $
symmetries. These global symmetries are not lost by gauge fixing, since the
original Lagrangian is both gauge invariant and globally symmetric.

It is interesting to show how the gauge fixed form of $\Theta ,X_{i}^{M}$
provide a basis for these global symmetries. Since the global symmetry tends
to change the gauge fixed form of $\Theta ,X_{i}^{M},$ there must be some
compensation from the gauge transformations kappa and Sp$\left( 2,R\right) $
to restore the gauge fixed forms. In particular let us examine the form of $%
\Theta $. The global and local supersymmetry transformations $\delta
_{\varepsilon +\kappa }\Theta $ given in the appendix simplify in this gauge
due to $\left( \tilde{\Theta}\Theta \right) _{a}^{\,\,\,b}=0$ 
\begin{equation}
\delta _{\varepsilon +\kappa }\Theta =\varepsilon +\not{X}_{i}\kappa
^{i}+\Theta \tilde{\varepsilon}\Theta +\frac{1}{2}\Theta \tilde{\Theta}\not%
{X}_{i}\kappa ^{i}.
\end{equation}
The global $\varepsilon $ has a lower component. But the lower component of $%
\Theta $ must be maintained at zero by the combined $\varepsilon $ and $%
\kappa $ transformations. Indeed this requirement is satisfied by taking $%
\left( \Gamma ^{+^{\prime }}\kappa ^{ia}\right) ^{\alpha }=0$ with 
\begin{eqnarray}
\varepsilon  &=&\left( 
\begin{array}{l}
\epsilon  \\ 
\lambda 
\end{array}
\right) \quad ,\quad \kappa ^{1}=\left( 
\begin{array}{l}
\lambda  \\ 
0
\end{array}
\right) ,\quad \kappa ^{2}=\left( 
\begin{array}{l}
k\left( \tau \right)  \\ 
0
\end{array}
\right)  \\
\kappa  &=&\not{X}_{i}\kappa ^{i}=\left( 
\begin{array}{l}
\not{p}k\left( \tau \right) +\not{x}\lambda  \\ 
-\lambda 
\end{array}
\right) ,\quad \delta _{\varepsilon +\kappa }\Theta =\left( 
\begin{array}{l}
\delta \theta  \\ 
0
\end{array}
\right)  \\
\delta \theta  &=&\epsilon +\not{p}k\left( \tau \right) +\not{x}\lambda
+\theta \left( \tilde{\lambda}\theta -\frac{1}{2}\,\tilde{\theta}\lambda
\right) .
\end{eqnarray}
where we have also used the gauge fixed form of $X_{i}^{M}$. The $-\lambda $
in the second line of $\kappa $ comes from $-X^{+^{\prime }}\Gamma
^{-^{\prime }}\kappa ^{1}$. The upper component $\epsilon $ is the standard
global SUSY parameter for the superparticle. Similarly, $k\left( \tau
\right) $ is the standard local kappa symmetry parameter. The lower
component of the global supersymmetry tends to add $\lambda $ to the lower
component of $\Theta $, shifting it away from the gauge fixed form, but by
taking the upper component of $\kappa ^{1}$ equal to the global $\lambda $
this is cancelled and the gauge fixed form of $\Theta $ is maintained. From
our fully covariant construction we know that $\lambda $ is the global
parameter of the special superconformal transformation. Therefore the
special superconformal transformation in some sense probes the hidden
fermionic dimensions that were gauge fixed to zero. This is similar to what
happens in the purely bosonic theory with special conformal transformations.
There the naive Lorentz boosts that mix $x^{\mu }$ with the extra dimensions 
$X^{\pm ^{\prime }}$ are compensated by the Sp$\left( 2,R\right) $ gauge
transformations. The combined Lorentz boost and Sp$\left( 2,R\right) $
transformation is the special conformal transformation, so that it may be
viewed as a hidden symmetry that probes the extra dimensions.

\section{Superalgebra generators in superparticle gauge}

The generators given by eqs. (\ref{psia})-(\ref{VA}) $\Psi ^{\alpha
a},J^{MN},T^A$ are gauge invariant under both the local Sp$\left( 2,R\right) 
$ and kappa supersymmetry transformations and therefore they are
observables. Together with the action, they can be evaluated in any fixed
gauge simply by inserting any gauge fixed form of $X_i^M,\Theta ^{\alpha a}$.

In the superparticle gauge the fermionic generators take the form (after
simplifications that follow from the series expansion of the square root and 
$\left( \tilde{\Theta}\Theta \right) _a^{\,\,\,b}=0$) 
\begin{eqnarray}
\Psi ^a &=&\frac 12L_{MN}\left( \Gamma ^{MN}\Theta -\frac 12\Theta \tilde{%
\Theta}\Gamma ^{MN}\Theta \right) ^a=\left( 
\begin{array}{l}
S^a \\ 
Q^a
\end{array}
\right) , \\
S^a &=&\not{x}\not{p}\theta ^a+\frac 12\theta ^b\left( \tilde{\theta}_b\not{p%
}\theta ^a\right) ,  \label{Sa} \\
Q^a &=&\sqrt{2}\left( \not{p}\theta ^a\right) ^\alpha .  \label{Qa}
\end{eqnarray}
where the interpretation of the upper/lower components are 
\begin{eqnarray}
S^a &=&special\,\,superconformal\,\,generator \\
Q^a &=&global\,\,SUSY\,\,generator\,\,\,in\,\,d\,\,dimensions
\end{eqnarray}
The bosonic generators $J^{MN},T^A$ in eq.(\ref{JMN},\ref{VA}) take the form
(dilatations $D$, translations $P^\mu $, special conformal transformations $%
K^\mu $, Lorentz transformations $J^{\mu \nu }$, internal symmetries $T^A$) 
\begin{eqnarray}
D &=&J^{+^{\prime }-^{\prime }}=L^{+^{\prime }-^{\prime }}+0,\quad \\
P^\mu &=&J^{+^{\prime }\mu }=L^{+^{\prime }\mu }+0 \\
K^\mu &=&J^{-^{\prime }\mu }=L^{-^{\prime }\mu }-\frac 12\left( \tilde{\theta%
}_a\gamma ^\mu \not{x}Q^a+h.c.\right) -\frac 1{2\sqrt{2}}\left( \tilde{\theta%
}_a\gamma ^\mu \theta ^b\right) \left( \tilde{\theta}_b\not{p}\theta
^a\right) \\
J^{\mu \nu } &=&L^{\mu \nu }+\frac 12\left( \tilde{\theta}_a\gamma ^{\mu \nu
}Q^a+h.c.\right) ,\quad \\
T^A &=&\tilde{\theta}_a\left( t^A\right) _{\,\,\,b}^aQ^b+h.c.
\end{eqnarray}
The explicit form of the purely bosonic part $L^{MN}$ is given in this gauge
by 
\begin{eqnarray}
L^{+^{\prime }\mu } &=&p^\mu ,\quad L^{\mu \nu }=x^\mu p^\nu -x^\nu p^\mu \\
L^{-^{\prime }\mu } &=&\frac 12x^2p^\mu -x\cdot px^\mu ,\quad L^{+^{\prime
}-^{\prime }}=x\cdot p,\quad
\end{eqnarray}
The fermionic part in $J^{+^{\prime }\mu }$ vanishes, thus yielding just the
momentum $J^{+^{\prime }\mu }=p^\mu $. In addition, the fermionic part in $D$
vanishes due to $\tilde{\theta}_a\gamma ^\mu \theta ^a=0$ for $d=3,4,6$.

The fermionic parts in $K^\mu ,J^{\mu \nu },T^A$ can be expressed in terms
of $\theta $ by replacing $Q^a,S^a$ from (\ref{Qa},\ref{Sa}). We then find
that for $d=3,4$ our results agree with the formulas in \cite{schwarz}
(3.20c, 3.20e for d=3) and ( 4.14c and 4.14g for d=4) after a Fierz
rearrangement. Our compact expressions for $J^{MN}$ and $T^A$ also agree
with the rest of the formulas (3.20 for d=3) and (4.14 for d=4) from \cite
{schwarz}. On the other hand, our gauge fixed expressions provide new
formulas for the $d=6$ superparticle. This result can also be obtained with
the methods of \cite{schwarz}\cite{private}.

\section{Discussion and future directions}

In this paper we have generalized the results of \cite{dualconf} to target
space supersymmetry. We have constructed an Sp($2,R$) gauge invariant
particle action which possesses manifest space-time SO(d,2) symmetry, global
supersymmetry and kappa supersymmetry. In particular, we have demonstrated
that the superconformal groups $OSp(N/4)$, $SU(2,2/N)$ and $OSp(8^{*}/N)$
familiar in the context of $AdS/CFT$ duality framework can be given unified
and explicit superparticle representations.

One way of gauge fixing produced the superparticle Lagrangian with
superconformal symmetry, thus showing that the results of \cite{schwarz}
appear only as a particular gauge in our framework. As pointed out in the
introduction there are other possible gauge choices for our Lagrangian,
which would provide explicit realizations of the various space-time
superconformal symmetries, such as superparticles moving in the background
of $AdS_D\times S^n$, supersymmetric H-atom, supersymmetric particle moving
in the black hole background, etc., to name a few. It may be interesting to
investigate the details of the theory in other gauges.

We think it would be very exciting to extend the same treatment for the case
of superstrings (some results in the same framework have been obtained for
various bosonic models \cite{toappear}). It would be particularly exciting
to see if various superstring theories can be indeed obtained as different
gauge choices of the same theory.

Another possible direction for exploration is to consider superconformal
algebras that do not have an obvious space-time interpretation, such as OSp$%
\left( 1/32\right) $ \cite{town}, OSp$\left( 16/2\right) $ \cite{mm}, OSp$%
\left( 1/32\right) \times $OSp$\left( 1/32\right) $ \cite{horava}, OSp$%
\left( 1/64\right) $ \cite{bars}, which have been argued from various points
of view to be important in M-theory or S-theory. Our method can be
generalized to construct Lagrangians that are invariant under these
supergroups. For example, for OSp$\left( 1/32\right) $ denote the generators
of OSp$\left( 1/32\right) $ as $L_{MN}$, $L_{M_1\cdots M_6}^{+}$ and $\Psi
_\alpha $ in the SO$(10,2)$ basis with parameters $l^{MN},$ $%
l_{+}^{M_1\cdots M_6}$ and $\Theta ^\alpha $ (32 component spinor). Consider
the coset element $t\left( l_{+},\Theta \right) $ in OSp$\left( 1/32\right)
/ $SO$\left( 10,2\right) $, where $l_{+}$ now represent extra bosonic
degrees of freedom, and then repeat the steps of our construction to obtain $%
\Omega ^{MN}\left( l_{+},\Theta \right) $ and a Lagrangian invariant under
global OSp$\left( 1/32\right) $ with transformations 
\begin{equation}
t\left( l_{+},\Theta \right) \rightarrow t\left( l_{+}^{\prime },\Theta
^{\prime }\right) =h^{-1}t\left( l_{+},\Theta \right) g\,\,.
\end{equation}
Here $g\in $OSp$\left( 1/32\right) $ global, and $h\left( l_{+},\Theta
;g\right) \in $SO$\left( 10,2\right) $ is the induced local Lorentz
transformation which also transforms $X_i^M\left( \tau \right) $. There is
also local symmetry that originates with transformations on the left side of 
$t\left( l_{+},\Theta \right) $ with fermionic as well as bosonic parameters 
$b_{+}^{M_1\cdots M_6}\left( \tau \right) ,\kappa ^\alpha \left( \tau
\right) $ that form a generalization of kappa supersymmetry. The
transformation is 
\begin{equation}
t\left( l_{+},\Theta \right) \rightarrow t\left( l_{+}^{^{\prime \prime
}},\Theta ^{^{\prime \prime }}\right) =h^{-1}t\left( b_{+},\kappa \right)
\,t\left( l_{+},\Theta \right)
\end{equation}
where again $h\left( b_{+},\kappa ;l_{+},\Theta \right) $ is the induced
local Lorentz transformation. The dynamical degrees of freedom in this
construction are $X_i^M\left( \tau \right) ,l_{+}^{M_1\cdots M_6}\left( \tau
\right) ,\Theta ^{\alpha}\left( \tau \right) $. The interpretation of the
extra bosonic degrees of freedom $l_{+}$ is not obvious from the space-time
point of view and we suspect that these are related to p-brane degrees of
freedom. In any case the Lagrangian thus constructed may be an interesting
toy model to explore these types of supersymmetries, involving two timelike
dimensions that are covariantly included as part of the SO(10,2) particle
position-momentum vectors $X^M\left( \tau \right) ,P^M\left( \tau \right) $.
In addition to OSp$\left( 1/32\right) $, the supergroups not included in
Nahm's classification thus become relevant. We hope to return to these
problems and give more details in the future \cite{future}.

\section{Appendix: some details of the construction}

Using (\ref{tmatrix}) the left-hand side and the right-hand side of (\ref
{ttht}) become respectively 
\begin{eqnarray}
&&t\left( \Theta _1\right) t\left( \Theta _2\right)  \nonumber \\
&=&\left( 
\begin{array}{cc}
\frac 1{\sqrt{1+\Theta _1\tilde{\Theta}_1}}\left( 1-\Theta _1\tilde{\Theta}%
_2\right) \frac 1{\sqrt{1+\Theta _2\tilde{\Theta}_2}} & \frac 1{\sqrt{%
1+\Theta _1\tilde{\Theta}_1}}\left( \Theta _1+\Theta _2\right) \frac 1{\sqrt{%
1+\tilde{\Theta}_2\Theta _2}} \\ 
-\frac 1{\sqrt{1+\tilde{\Theta}_1\Theta _1}}\left( \tilde{\Theta}_1+\tilde{%
\Theta}_2\right) \frac 1{\sqrt{1+\Theta _2\tilde{\Theta}_2}} & \frac 1{\sqrt{%
1+\tilde{\Theta}_1\Theta _1}}\left( 1-\tilde{\Theta}_1\Theta _2\right) \frac
1{\sqrt{1+\tilde{\Theta}_2\Theta _2}}
\end{array}
\right)
\end{eqnarray}
and 
\begin{eqnarray}
&&h\left( \Theta _{1,}\Theta _2\right) t\left( \Theta _{12}\right)  \nonumber
\\
&=&\left( 
\begin{array}{ll}
h_{12} & 0 \\ 
0 & \tilde{h}_{12}
\end{array}
\right) \left( 
\begin{array}{ll}
\frac 1{\sqrt{1+\Theta _{12}\tilde{\Theta}_{12}}} & \Theta _{12}\frac 1{%
\sqrt{1+\tilde{\Theta}_{12}\Theta _{12}}} \\ 
-\frac 1{\sqrt{1+\tilde{\Theta}_{12}\Theta _{12}}}\tilde{\Theta}_{12} & 
\frac 1{\sqrt{1+\tilde{\Theta}_{12}\Theta _{12}}}
\end{array}
\right)
\end{eqnarray}
By comparing these two equations we infer

\begin{eqnarray}
\Theta _{12} &=&\sqrt{1+\Theta _2\tilde{\Theta}_2}\left( 1-\Theta _1\tilde{%
\Theta}_2\right) ^{-1}\left( \Theta _1+\Theta _2\right) \frac 1{\sqrt{1+%
\tilde{\Theta}_2\Theta _2}} \\
h_{12} &=&\frac 1{\sqrt{1+\Theta _1\tilde{\Theta}_1}}\left( 1-\Theta _1%
\tilde{\Theta}_2\right) \frac 1{\sqrt{1+\Theta _2\tilde{\Theta}_2}}\sqrt{%
1+\Theta _{12}\tilde{\Theta}_{12}} \\
\tilde{h}_{12} &=&\frac 1{\sqrt{1+\tilde{\Theta}_1\Theta _1}}\left( 1-\tilde{%
\Theta}_1\Theta _2\right) \frac 1{\sqrt{1+\tilde{\Theta}_2\Theta _2}}\sqrt{1+%
\tilde{\Theta}_{12}\Theta _{12}}
\end{eqnarray}
Then we can specialize to the infinitesimal transformations ($\Theta
_1=\Theta $, $\Theta _2=\varepsilon $, and small) and find 
\begin{equation}
\delta _\varepsilon \Theta \approx \varepsilon +\Theta \tilde{\varepsilon}%
\Theta ,\quad
\end{equation}
If we set ($\Theta _2=\Theta $, $\Theta _1=\kappa $, and small) we get

\begin{equation}
\delta _\kappa \Theta =\sqrt{1+\Theta \tilde{\Theta}}\,\kappa \,\sqrt{1+%
\tilde{\Theta}\Theta }\,.
\end{equation}


\begin{thebibliography}{99}
\bibitem{dualconf}  I. Bars, C. Deliduman and O. Andreev, Phys. Rev. {\bf D58%
} (1998) 066004, or hep-th/9803188.

\bibitem{dualH}  I. Bars, Phys. Rev. {\bf D58 (}1998) 066006, or
hep-th/9804028.

\bibitem{dualsusy}  I. Bars and C. Deliduman, Phys. Rev. {\bf D58} (1998)
106004, or hep-th/9806085.

\bibitem{dualicmp}  I. Bars, ``Two-Time Physics'', hep-th/9809034, to appear
in the proceedings of the XXIII$^{rd}$ Int. Colloq. on Group Theor. Methods
in Physics, Hobart, Australia, July 1998, hep-th/9809034.

\bibitem{lifting}  I. Bars, ``Hidden Symmetries, AdS$_D\times $S$^n$, and
the lifting of one-time physics to two-time physics'', hep-th/9810025, to
appear in Phys. Rev. D.

\bibitem{maldacena}  J. Maldacena, hep-th/9711200; S. S. Gubser, I. R.
Klebanov and A. M. Polyakov, hep-th/9802109; E. Witten, hep-th/9802150.

\bibitem{maldstrom}  J. D. Brown and M. Henneaux, Comm. Math. Phys. ({\bf 104%
} (1986) 207; A. Strominger, JHEP {\bf 2} (1998) 9; J. Maldacena and A.
Strominger, hep-th/9804085; J. de Boer, hep-th/9806104.

\bibitem{multi}  I. Bars and C. Deliduman, Phys. Lett. {\bf B417 }(1998) 24,
or hep-th/9710066; E. Sezgin and Rudychev, hep-th/9711128; I. Bars and C.
Kounnas, Phys. Lett. {\bf B402} (1997) 25; Phys. Rev. {\bf D56} (1997) 3664.

\bibitem{schwarz}  J. H. Schwarz, Nucl. Phys. {\bf B185} (1981) 221.

\bibitem{marnelius}  R. Marnelius, Phys. Rev. {\bf D20 }(1979){\bf \ }2091.

\bibitem{siegel}  W. Siegel, Int. J. Mod. Phys. {\bf A3 }(1988){\bf \ }2713.

\bibitem{barsgunay}  I. Bars and M. G\"{u}naydin, Phys. Rev. {\bf D22}
(1980) 1403.

\bibitem{sorokin}  D. Sorokin, V.I. Tkach and D.V. Volkov, Mod. Phys. Lett.
A4 (1989) 901.

\bibitem{tonin}  I. Bandos, P. Pasti, D. Sorokin and M. Tonin,
hep-th/9705064.

\bibitem{howe}  P. Howe, E. Sezgin and P. West, hep-th/9705093.

\bibitem{mm}  M. G\"{u}naydin, D. Minic, Nucl. Phys. {\bf B523} (1998) 145,
hep-th/9802047; M. G\"{u}naydin, hep-th/9803138.

\bibitem{private}  J. Schwarz, private communication.

\bibitem{town}  I. Bars, hep-th/9608061; P. K. Townsend, hep-th/9708034; D.
Sorokin and P.K. Townsend, hep-th/9709007.

\bibitem{horava}  P. Horava, hep-th/9712130.

\bibitem{bars}  I. Bars, Phys. Rev. {\bf D55} (1997) 2373 or hep-th/9607112;
Phys. Lett. {\bf B403} (1997) 257 or hep-th/9704054.

\bibitem{toappear}  I. Bars, C. Deliduman and D. Minic, to appear.

\bibitem{future}  I. Bars, C. Deliduman and D. Minic, in preparation.
\end{thebibliography}
\end{document}